\documentclass[%
 reprint,
superscriptaddress,
 amsmath,amssymb,
 aps,
prx,
floatfix,
]{revtex4-2}

\usepackage{graphicx}
\usepackage{dcolumn}
\usepackage{bm}
\usepackage{cancel}%
\usepackage{xcolor}
\usepackage[normalem]{ulem}

\newcommand{\mydef}[1]{\vspace{0.1cm} \noindent \textbf{Definition:} #1\\}

\newcommand{\prop}[1]{\vspace{0.1cm} \noindent \textbf{Proposition:} #1\\}
\newcommand{\RNum}[1]{\uppercase\expandafter{\romannumeral #1\relax}}

\begin{document}

\title{Energetic rigidity \RNum{1}.\\
A unifying theory of mechanical stability}%

\author{Ojan Khatib Damavandi}
\affiliation{%
 Department of Physics, Syracuse University, Syracuse, New York 13244, USA}%
\author{Varda F.\ Hagh}
\affiliation{%
 James Franck Institute, University of Chicago, Chicago, IL 60637, USA}%
\author{Christian D.\ Santangelo}
\email[To whom correspondence should be addressed:\\]{cdsantan@syr.edu}
\affiliation{%
 Department of Physics, Syracuse University, Syracuse, New York 13244, USA}
\author{M.\ Lisa Manning}
\email[To whom correspondence should be addressed:\\]{mmanning@syr.edu}
\affiliation{%
 Department of Physics, Syracuse University, Syracuse, New York 13244, USA}

\begin{abstract}
Rigidity regulates the integrity and function of many physical and biological systems. This is the first of two papers on the origin of rigidity, wherein we propose that ``energetic rigidity,'' in which all non-trivial deformations raise the energy of a structure, is a more useful notion of rigidity in practice than two more commonly used rigidity tests: Maxwell-Calladine constraint counting (first-order rigidity) and second-order rigidity. We find that constraint counting robustly predicts energetic rigidity only when the system has no states of self stress. When the system has states of self stress, we show that second-order rigidity can imply energetic rigidity in systems that are not considered rigid based on constraint counting, and is even more reliable than shear modulus. We also show that there may be systems for which neither first nor second-order rigidity imply energetic rigidity. The formalism of energetic rigidity unifies our understanding of mechanical stability and also suggests new avenues for material design. 
\end{abstract}


\maketitle

\section*{Introduction}\label{sec:intro}
How do we know if a material or structure is rigid? If we are holding it in our hands, we might choose to push on it to determine whether an applied displacement generates a proportional restoring force. If so, we say it is rigid. A structure that does not push back, on the other hand, would be said to be floppy. In this paper, we call this intuitive definition of rigidity ``energetic rigidity'' by virtue of the fact that small deformations increase the elastic energy of the structure. In many situations of interest, it is impossible or impractical to push on a structure to measure the restoring force. In designing new mechanical metamaterials, for example, we would like to sort through possible designs quickly, without having to push on every variation of a structure. In biological tissues such as the cartilage of joints or the bodies of developing organisms, it is often difficult to develop non-disruptive experimental rheological tools at the required scale. Or we may wish to understand how some tissues can tune their mechanical rigidity in order to adapt such functionality into new bio-inspired materials. To that end, we would like a theory that can predict whether a given structure is energetically rigid rapidly and without the need for large-scale simulations or experiments.

This has inspired the search for proxies: simple tests that, when satisfied, imply a structure is energetically rigid \cite{Maxwell1864, Calladine1978, Calladine1991, Connelly1996, Connelly2015}. The standard (and first) proxy for rigidity in particulate systems comes from Maxwell \cite{Maxwell1864}. When two particles interact, for example through a contact, that interaction constrains each particle's motion. ``Structural rigidity'' refers to whether those interaction constraints prevent motion in the system. If a system has fewer constraints than the particles have degrees of freedom, it is said to be underconstrained and therefore one expects it to be floppy. In contrast, overconstrained systems are said to be ``first-order rigid." This thinking has been successfully applied to many examples of athermal systems, such as jammed granular packings, randomly diluted spring networks, and stress diluted networks \cite{Lechenault2008, VanHecke2009, Ellenbroek2015, Liarte2019}. A straightforward extension of Maxwell's argument, known as the Maxwell-Calladine index theorem~\cite{Calladine1978,Lubensky2015}, shows that one should also subtract the number of states of self stress, equilibrium states of the system that can carry a load, because they arise from redundant constraints. In hinge-bar networks, these ideas can be exploited to design mechanical metamaterials with topologically protected mechanisms~\cite{Lubensky2015, Rocklin2016, Bertoldi2017, Mao2018, Zhou2018}.

Yet, this thinking is certainly wrong in general. It is well-known that underconstrained spring networks can be rigidified if put under enough strain~\cite{Tang1988, Storm2005, Wyart2008, Huisman2011,Sheinman2012, Feng2016, Sharma2016, Vermeulen2017,Rens2018,Merkel2019,Shivers2019}. And there are special configurations of even unstressed networks, \text{e.g.} colinear springs pinned down at both ends or honeycomb lattice in a periodic box~\cite{Rens2016}, which are rigid despite being under-coordinated. That this occurs because of nonlinear effects has already been highlighted by mathematicians and engineers in the context of the bar-joint frameworks, origami, and tensegrities \cite{Calladine1991, Connelly1996, Connelly2015, Guest2006, Chen2018, Holmes-Cerfon2020}. In particular, Connelly and Whitely~\cite{Connelly1996} demonstrate that there may exist states where a different proxy, termed ``second-order rigidity", is sufficient to ensure that the constraints are preserved. Because of these nonlinear effects, determining whether even a planar network of springs is rigid is NP-hard \cite{Abbott} and, consequently, there is no simple theory that can determine if a mechanical system is truly rigid. Maxwell constraint counting works because these non-generic configurations are ostensibly rare.

In many physical systems of interest, however, the dynamics or boundary conditions drive the system towards specific, non-generic states~\cite{Sartor2020}. These non-generic states can behave differently than we would expect from rigidity proxies. For example, even in overconstrained elastic networks, prestresses have been shown to affect the stability of the system~\cite{Bose2019}.  As another example, deformable particles with bending constraints have been observed to jam at a hyperstatic point~\cite{Treado2021}. Therefore, instead of demonstrating the existence of states that are first-order or second-order (and thus structurally) rigid, we instead ask a different question: what can we say about energetic rigidity for systems that are at an energy minimum and correspond to highly non-generic states selected by physical dynamics? In particular, is it possible to find or design structures where motions preserve the energy but not the individual constraints? In an important sense, such a structure would still be floppy.

To answer this question we develop a generalized formalism for understanding the rigidity of energetically stable physical materials. Specifically, we demonstrate that the onset of rigidity upon tuning a continuous parameter emerges from the effects of geometric incompatibility arising from higher-order corrections to Maxwell-Calladine constraint counting. Depending on the prestresses in the system and features of the eigenvalue spectrum, we identify different cases where first-order or second-order rigidity imply energetic rigidity. We also demonstrate cases where second-order rigidity is a more reliable proxy for energetic rigidity than even the shear modulus, the standard measure of rigidity used in physics.

\section{Formalism}\label{sec:formalism}
In this section, we will introduce notation and summarize some of the standard proxies of rigidity and structural rigidity that arise in physics and mathematics. We assume the state of the system is described by $N_{dof}$ generalized coordinates, $x_n$. For example, the coordinates $\{ x_n \}$ might represent the components of the positions of all vertices in a spring network. We also introduce $M$ strains of the form $f_\alpha( \{x_n\} )$ and assume the physical system is characterized by the Hooke-like energy, $E$, of the form
\begin{equation}\label{eq:energy}
    E = \frac{1}{2} \sum_{\alpha=1}^M k_\alpha f_\alpha(\{x_n\})^2,
\end{equation}
where  $k_\alpha > 0$ is the stiffness associated with each strain.
Since the strain functionals $f_\alpha( \{x_n\} )$ are in principle general, energies of the form of Eq.~(\ref{eq:energy}) encompass a broad array of physical systems with Hookean elasticity.

As a concrete example, for a $d-$dimensional spring network of $N$ vertices connected via $M$ springs with rest length $L_0$ in a fixed periodic box, $N_{dof}= d N$ and the strain associated with spring $\alpha$ connecting vertices $i$ and $j$ at positions $\mathbf{X}_i$ and $\mathbf{X}_j$ is simply the strain of the spring, $f_\alpha = L_\alpha - L_0$, where $L_\alpha = \lvert \mathbf{X}_i - \mathbf{X}_j\lvert$ is the actual length of the spring. Without loss of generality, we absorb $k_\alpha$ into $f_\alpha$ by re-scaling it by $\sqrt{k_\alpha}$ and writing $E = \sum_{\alpha=1}^M f_\alpha^2/2$.

We can capture the intuitive notion of rigidity or floppiness by considering the behavior of Eq.~(\ref{eq:energy}) under deformations. A system is \emph{energetically rigid} if any \emph{global} motion that is not a trivial translation or rotation increases the energy. A global motion is one that extends through the entire system so as to exclude rattlers or danglers. If there exists a nontrivial, global motion that preserves the energy, we call the system \emph{floppy}. If, for a given system at an energy minimum, all the strains vanish, $f_\alpha = 0$ for all $\alpha$, and the system is \emph{unstressed}. Otherwise, we say the system is \emph{prestressed}.

The relationship between structural and energetic rigidity arises when we treat the generalized strains, $f_\alpha$, as the constraints in Maxwell-Calladine counting arguments. However, while structural rigidity depends on geometry only, we will see that energetic rigidity must depend on the particular energy functional. Nevertheless, it is natural that a useful definition of floppiness would depend on the energy functional itself.

\begin{table*}[t]
    \centering
    \begin{tabular}{l|l}
        A system is \dots & when \dots \\
        \hline
        Energetically rigid & any nontrivial global motion increases the energy\\
        \hline
        Structurally rigid & no nontrivial global motion preserves the constraints $f_\alpha$\\
        \hline
        First-order rigid & no nontrivial global motion preserves the constraints $f_\alpha$ to first order\\
        \hline
        Second-order rigid & no nontrivial global motion preserves the constraints $f_\alpha$ to second order
    \end{tabular}
    \caption{Different definitions of rigidity.} 
    \label{tab:my_label}
\end{table*}

\subsection{Standard proxies of energetic rigidity}

Experimentally, the standard proxy used to determine whether the system is energetically rigid is the shear modulus, $G$, defined as the second derivative of energy with respect to a shear variable $\gamma$ in the limit of zero shear~\cite{Merkel2018,Wang2020}:
\begin{align}
    G &= \frac{1}{V}\frac{\mathrm{d}^2E}{\mathrm{d}\gamma^2}\nonumber\\
    &=\frac{1}{V}\left(\frac{\partial^2E}{\partial \gamma^2}-\sum_l\frac{1}{\lambda_l}\left[\sum_n\frac{\partial^2E}{\partial \gamma \partial x_n}u^{(l)}_n\right]\right),
    \label{eq:shearModulus}
\end{align}
where $V$ is the volume of the system while $\lambda_l$ and $u^{(l)}_n$ are respectively the eigenvalues and eigenvectors of the Hessian matrix, $H_{nm}=\partial^2 E/\partial x_n \partial x_m$, and the sum excludes eigenmodes with $\lambda_l=0$.
When $G \ne 0$, the system is certainly energetically rigid. Note that this is closely allied with the mathematical notion of prestress stability \cite{Connelly1996} (see Appendix~\ref{app:formalism}). On the other hand, if $H_{nm}$ has global, nontrivial zero eigenmodes (or more precisely, zero eigenmodes that overlap with the shear degree of freedom), $G=0$~\cite{Merkel2018}.

Importantly, defining rigidity based on $G$ is \emph{not equivalent} to energetic rigidity. Specifically, $G \ne 0$ implies the system is energetically rigid, but $G=0$ does not imply floppiness. As highlighted in Appendix~\ref{app:formalism} there may be quartic corrections in $\delta x_n$ that increase the energy even with vanishing shear modulus. Moreover, in many cases of interest these quartic corrections are expected to dominate precisely at the onset of rigidity.

A definition of rigidity based on $G$ is equivalent to examining the Hessian matrix $H$ directly: if $H$ is positive definite on the global, non-trivial deformations, then the system is also energetically rigid. Writing out the Hessian matrix in terms of the constraints, we find
\begin{align}
    H_{nm} &= \frac{\partial^2 E}{\partial x_n \partial x_m} = \sum_\alpha  \left[\frac{\partial f_\alpha}{\partial x_n}\frac{\partial f_\alpha}{\partial x_m} + f_\alpha\frac{\partial^2 f_\alpha}{\partial x_n\partial x_m} \right]\nonumber \\
    &=(R^T R)_{nm} + \mathcal{P}_{nm},
    \label{eq:Hessian}
\end{align}
where 
\begin{equation}
    R_{\alpha n } = \frac{\partial f_\alpha}{\partial x_n}
    \label{eq:rigiditymatrix}
\end{equation}
is known as the rigidity matrix. We call $(R^T R)_{nm}$ the Gram term (as it is the Gramian of rigidity matrix), and $\mathcal{P}_{nm}$ the prestress matrix because it is only non-zero if $f_\alpha \neq0$ (Gram term and prestress matrix are sometimes called stiffness matrix and geometric stiffness matrix respectively in structural engineering~\cite{Connelly1996,Guest2006}).
If the Hessian has at least one global nontrivial zero direction, we obtain the necessary (but not sufficient) condition for floppiness,
\begin{align}
\sum_{nm} \mathcal{P}_{nm} \delta x_n \delta x_m &= -\sum_{nm} (R^T R)_{nm}\delta x_n \delta x_m\nonumber\\
&= -\sum_\alpha   \left(\sum_n\frac{\partial f_\alpha}{\partial x_n} \delta x_n\right)^2,
\label{eq:floppiness-condition}
\end{align}
where the sum over $\alpha$ is over all constraints and, again, trivial Euclidean modes have been excluded. Analogous to our discussion of $G$ above, a definition of rigidity based on $H$ is also not equivalent to energetic rigidity, due to the importance of quartic terms in cases of interest (including at the transition point).

\subsection{Proxies of structural rigidity: the first- and second-order rigidity tests}
The existence of any global, non-trivial, and continuous motion of the system $x_n(t)$ that preserves the constraints $f_\alpha( \{x_n(t)\} )$ implies the system is floppy. A system is structurally rigid when no such motions exist, a definition highlighted in Table~\ref{tab:my_label}. Energetic rigidity is not necessarily equivalent to structural rigidity when the system is prestressed ($E>0$), though the two are the same when $E = 0$, as discussed in more detail later.

Though determining whether a system is structurally rigid is NP-hard \cite{Abbott}, there are several simpler conditions that, if they hold true, imply that a system is structurally rigid \cite{Calladine1978, Calladine1991, Connelly1996, Connelly2015}. These tests, and in particular the first- and second-order rigidity tests, are reviewed in more detail in Appendix~\ref{app:formalism} and briefly summarized in Table \ref{tab:my_label}.

The first-order rigidity test arises by considering first-order perturbations to the constraints, $\delta f_\alpha = \sum_n \partial f_\alpha/\partial x_n \delta x_n$. We define a linear (first-order) zero mode (LZM) $\delta x_n^{(0)}$ as one that preserves $f_\alpha$ to linear order,
\begin{equation}
    \sum_n \frac{\partial f_\alpha}{\partial x_n}  \delta x_n^{(0)} = \sum_n R_{\alpha n } { \delta x_n^{(0)}} = 0.
    \label{eq:ZM}
\end{equation}
We can see that LZMs are in the right nullspace of the rigidity matrix. Excluding Euclidean motions, a nontrivial LZM is often called floppy mode (FM) in physics \cite{Lubensky2015}.  A system with no nontrivial LZM is \emph{first-order rigid} and, indeed, in such systems first-order rigidity implies structural rigidity as defined in Table \ref{tab:my_label} \cite{Calladine1991, Connelly1996}.

Maxwell constraint counting suggests that an overconstrained system ($N_{dof}<M$) must be rigid while an underconstrained system ($N_{dof}>M$) must be floppy. If $R_{\alpha n}$ is full rank for a domain of configurations, this intuition is assuredly true. Yet, there are examples of contrivances that appear overconstrained yet move \cite{schicho2021and}, as well as underconstrained systems that are rigid.

When an underconstrained system is rigid, it must be in configurations for which $R_{\alpha n}$ fails to be full rank. Thus, the system must exhibit a state of self stress, defined as a vector $\sigma_{\alpha}$ in the left nullspace of the rigidity matrix:
\begin{equation}
    \sum_{\alpha} \sigma_{\alpha} R_{\alpha n} = 0.
    \label{eq:SSS}
\end{equation}
The Maxwell-Calladine index theorem (also known as the rigidity rank-nullity theorem) states that $N_{dof}-M = N_0 - N_s$, where $N_0$ is the number of LZMs and $N_s$ is the number of states of self stress~\cite{Calladine1978}.

To understand this case, we study motions that preserve $f_\alpha$ to second order in $\delta x_n$. Taylor expansion of $f_\alpha$ results in:
\begin{equation}
    \delta f_\alpha \approx \sum_n R_{\alpha n} \delta {x}_n  +\frac{1}{2} \sum_{nm} \frac{\partial^2 f_\alpha}{\partial x_n\partial x_m} \delta x_n \delta x_m=0,
    \label{eq:deltaf}
\end{equation}
where we used Eq.~(\ref{eq:rigiditymatrix}) for the linear term in the expansion. If the only LZMs that satisfy Eq.~(\ref{eq:deltaf}) are trivial ones, the system is called \emph{second-order rigid} and, consequently, is structurally rigid \cite{Calladine1991,Connelly1996}. It can be shown that a LZM, $ \delta x_n^{(0)}$, must satisfy
\begin{eqnarray}\label{eq:secondorder}
    \sum_\alpha \sum_{nm} \sigma_{\alpha, I} \frac{\partial^2 f_\alpha}{\partial x_n\partial x_m}  \delta x_n^{(0)} \delta x_m^{(0)} = 0,
\end{eqnarray}
for all states of self stress $\sigma_{\alpha,I}$ and solutions to Eq.~(\ref{eq:SSS}) to be a second-order zero mode (\cite{Connelly1996,Connelly2015}; Appendix~\ref{app:formalism}).

Testing for second-order rigidity is not always easy,  particularly when there are more than one states of self stress~\cite{Holmes-Cerfon2020}. Thus, it is useful to define a stronger rigidity condition called \emph{prestress stability} which looks for a single self stress, $\sigma_{\alpha,I}$ for which Eq.~(\ref{eq:secondorder}) has no solution~\cite{Connelly1996}.  If such a self stress exists, the system is said to be \textit{prestress stable}, and in the case of underconstrained systems it is second-order rigid as well. Note that the inverse is not always true, \textit{i.e.}, second-order rigidity does not imply prestress stability: for a second-order rigid system with more than one self stress, individual FMs could still satisfy Eq.~(\ref{eq:secondorder}) for some self stresses, but there is not a self stress for which all FMs satisfy Eq.~(\ref{eq:secondorder}).
Connelly and Whitely have shown, however, that a system that is first-order rigid is also prestress stable \cite{Connelly1996}.

Finally, we note that going beyond second order is less helpful than one might suppose. There are examples of systems that are rigid only at third order or beyond yet remain floppy \cite{Connelly1994}.

\subsection{How common are non-generic states?}
As we have seen, being able to use Maxwell constraint counting as a proxy for rigidity relies on being in a generic configuration. One might suppose that such cases must be rare but, in fact, non-generic configurations seem to arise physically quite often. Consider the Euler-Lagrange equations for a system with the energy of Eq.~(\ref{eq:energy}) at an extremum,
\begin{equation}
    \sum_\alpha   f_\alpha \frac{\partial f_\alpha}{\partial x_n} = \sum_\alpha   f_\alpha R_{\alpha n} = 0, \ \forall n.
    \label{eq:euler-lagrange}
\end{equation}
For a system that is not prestressed, $f_\alpha = 0$ and the above equation is satisfied trivially. For a system that is prestressed, $f_\alpha \neq 0$, $f_\alpha$ must be a state of self stress. Note, however, the converse is not true. The existence of states of self stress only depends on the geometry of the system and does not imply that the system has to be prestressed. For example, take a system with constraints $f_\alpha(\{x_n\}) = \mathcal{F}_\alpha(\{x_n\}) - F_\alpha$ at a particular mechanically stable configuration $\{\bar{x}_n\}$ that has a state of self stress and choose $F_\alpha = \mathcal{F}_\alpha(\{\bar{x}_n\})$. The system will be unstressed at $\{\bar{x}_n\}$ but still has a state of self stress. An example is the honeycomb lattice in a periodic boundary condition where all edge rest lengths are set to be equal to the actual edge lengths.

Thus if we put a system under an external tension so that it is unable to find a stress-free configuration under energy minimization, it will naturally evolve to a non-generic configuration having at least one self stress. In these cases, it would be surprising for Maxwell constraint counting to work; then the relationship between energetic and structural rigidity becomes more complex.

\section{Relating structural rigidity to energetic rigidity}
If a system is structurally rigid, can we also say it is energetically rigid? More specifically, when do the proxies of structural rigidity actually imply energetic rigidity? The number of self stresses, it turns out, can be used to classify the relationship between structural and energetic rigidity.

\subsection*{Case 1: The system has no self stresses ($N_s = 0$)}
When a system has no self stresses, first-order rigidity -- \textit{i.e.}, constraint counting -- is a good proxy for energetic rigidity. Since there are no self stresses, Eq.~(\ref{eq:euler-lagrange}) implies that the system is also unstressed, and Eq.~(\ref{eq:floppiness-condition}) reduces to
\begin{equation}
    \sum_\alpha   (\sum_n \partial_n f_\alpha \delta x_n)^2=0.
    \label{eq:unstressed-rigidity}
\end{equation}
The solutions are LZMs, $ \delta x_n^{(0)}$ (Eq.~(\ref{eq:ZM})). If a system does not have any FMs, it is energetically rigid. An energetically rigid system with no states of self stress is also called isostatic. This also means that there are no motions that preserve $f_\alpha$ even to first order, thus the system is first-order rigid. 
Examples of systems belonging to Case 1 include underconstrained and unstressed spring networks, unstressed vertex models with no area terms, and the special, non-generic frames described in Figs.~4(a)-(c) of \cite{Lubensky2015}.

\subsection*{Case 2: The system has at least one self stress ($N_s  \geq 1$)}
Once a system has a self stress, the relationship between energetic rigidity and structural rigidity becomes more subtle. Even a system that is first-order rigid may not be energetically rigid under some conditions.  For instance, jammed packings of soft particles are first-order rigid. However, in these packings, one can increase the prestress forces (for example by multiplying all the contact forces by a constant value as is shown in~\cite{hagh2021transient}) and push the lowest non-trivial eigenvalue of the Hessian to zero without leading to any particle rearrangements. In this case, the system is first-order rigid but not necessarily energetically rigid, and thus first-order rigidity does not always imply energetic rigidity (Fig.~\ref{fig:flowchart}).

An underconstrained system may also be structurally rigid but not necessarily energetically rigid. For example, consider an underconstrained system that is prestress stable for self stress $\sigma_{\alpha,1}$. Choose a prestress along this self stress, $\tilde{f}_\alpha = c \sigma_{\alpha,1}$ for some $c>0$ which defines an energy functional $\widetilde{E} = \sum_\alpha \tilde{f}_\alpha^2/2$. It follows from the assumption of prestress stability that the prestress matrix $\widetilde{\mathcal{P}}_{nm}$ defined for $\widetilde{E}$ is positive definite on the space of FMs. Therefore, if the actual energy of the system $E = \widetilde{E}$, $H_{nm}$ would be positive definite and the system energetically rigid at quadratic order.

However, $E = \widetilde{E}$ is only guaranteed if the system is prestressed along a unique state of self stress. For example, one can imagine a prestress stable system with more than one self stress that is driven to $f_\alpha = \sum_{I} c_I \sigma_{\alpha,I}$ by the dynamics such that $H_{nm}$ is not positive definite. Conversely, only if the system is energetically rigid at quadratic order, it is guaranteed to be prestress stable. For instance, a system may be energetically rigid at quartic order, which is the case for underconstrained systems at the critical point of rigidity transition as we will see later; such a system is second-order rigid (Appendix A) but not necessarily prestress stable.

We now ask the question: when does first- or second-order rigidity imply energetic rigidity? We identify two cases (Case 2A and 2B), which encompass several examples of physical interest, where both first-order and second-order rigidity imply energetic rigidity, and demonstrate that second-order rigidity is a better proxy for energetic rigidity than the shear modulus. We identify a third case (Case 2C) where neither first- or second-order rigidity imply energetic rigidity -- for example there may be systems with large prestresses that do not preserve $f_\alpha$ to second-order but preserve energy. We classify these distinct cases using the eigenspectrum of $\mathcal{P}_{nm}$ and the states of self stress. In all the cases, we will assume that if the system has FMs, at least one is global.

\subsubsection*{Case 2A: The system is unstressed ($\mathcal{P}_{nm} = 0$)}
This case includes systems with either no prestress, $f_\alpha =0$, or systems for which the prestress is perpendicular to its second-order expansion such that $\mathcal{P}_{nm} = \sum_\alpha   f_\alpha \partial_n \partial_m f_\alpha =0$. If the system is first-order rigid, it is again energetically rigid. If there are global FMs, $G=0$; however, it can be shown (Appendix~\ref{app:formalism}) that the fourth order expansion of energy for these modes will be 
\begin{equation}\label{eq:E_fourth}
    \delta E \approx \frac{1}{8} \sum_{I=1}^{N_s} \left[\sum_{\alpha,nm}   \sigma_{\alpha, I}\ \partial_n \partial_m f_\alpha\ \delta x_n^{(0)} \delta x_m^{(0)}\right]^2
\end{equation}
Therefore, if the system is second-order rigid in the space of its global FMs, it is energetically rigid even though $G=0$. Examples include random regular spring networks with coordination number $z=3$ and vertex models exactly at the rigidity transition.

\subsubsection*{Case 2B: $\mathcal{P}_{nm}$ is positive semi-definite}
For a system with a positive semi-definite $\mathcal{P}_{nm}$, the Hessian has a zero eigenmode if and only if both LHS and RHS of Eq.~(\ref{eq:floppiness-condition}) are zero for $\delta x_n$. The RHS is zero only for LZMs. Then if the system is first-order rigid, it is again energetically rigid. For a system with global FMs, we reduce Eq.~(\ref{eq:floppiness-condition}) to
\begin{equation}
    \sum_{nm} \mathcal{P}_{nm} \delta x_n^{(0)} \delta x_m^{(0)} = \sum_{nm} \sum_\alpha    f_\alpha \partial_n \partial_m f_\alpha \delta x_n^{(0)} \delta x_m^{(0)} = 0,
    \label{eq:prestressed-rigidity}
\end{equation}
where $x_n^{(0)}$ is now a global FM. We show below that second-order rigidity implies energetic rigidity, but depending on $N_s$, $G$ may be zero.

\textbf{If the system has a single self stress}: Calling this state of self stress $\sigma_\alpha$, we conclude from Eq.~(\ref{eq:euler-lagrange}) that $f_\alpha \propto \sigma_\alpha$, meaning Eq.~(\ref{eq:prestressed-rigidity}) is identical to Eq.~(\ref{eq:secondorder}) in this case. This means that if this system is second-order rigid, it is energetically rigid and $G>0$. We demonstrate in a companion paper~\cite{damavandi2021b} that both spring networks under tension and vertex models with only the perimeter term fall into this category. 

\textbf{If the system has multiple self stresses}: In Appendix~\ref{app:formalism} we show that if the system is second-order rigid in the space of global FMs, it is energetically rigid (Eq.~(\ref{eq:E_fourth})). However, the Hessian may still have zero eigenmodes if in the minimized state $f_\alpha$ is a linear combination of self stresses that satisfies Eq.~(\ref{eq:prestressed-rigidity}). This suggests that the system may be energetically rigid but with $G=0$. We have not been able to identify an example of a second-order rigid system with multiple self stresses and $G=0$, but if one exists, it may lead to interesting ideas for material design.

\subsubsection*{Case 2C: $\mathcal{P}_{nm}$ has negative eigenvalues}
In this case, we have been unable to derive analytic results for whether first-order or second-order rigidity implies energetic rigidity. As the models that fall into this class are quite diverse, it is likely that more restrictive conditions are necessary in specific cases to develop analytic results.

One example in this category is vertex models with an area term in addition to a perimeter term when prestressed. In the companion paper~\cite{damavandi2021b}, we demonstrate numerically that in such models there is always only one state of self stress that is non-trivial, and that $\mathcal{P}_{nm}$ has negative eigenvalues. However, the Hessian itself is still positive-definite (excluding trivial LZMs) and therefore the system is energetically rigid.  Another example is a rigid jammed packing, which exhibits quite different behavior for the eigenspectra of $\mathcal{P}_{nm}$.

More generally, we cannot rule out the possibility that there may be examples where the Hessian of a first-order or second-order rigid system could have global zero directions for non-zero modes. Such a system would be marginally stable because if any negative eigenmode of $\mathcal{P}_{nm}$ becomes too negative, the Hessian would have a negative direction and the system would not be at an energy minimum anymore. Furthermore, states of self stress place the same constraints as in Eq.~(\ref{eq:secondorder}) on these non-zero modes.  If those constraints are not satisfied, the energy would increase at fourth order (Appendix~\ref{app:formalism}), suggesting that again the shear modulus could be zero while the energy is not preserved. Even though it is highly non-generic, this case could aid in the design of structures that become unstable by varying the prestress~\cite{Bose2019} or new materials that are flexible even though individual constraints are not preserved. 

\begin{figure}[htp]
    \centering
    \includegraphics[width=\linewidth]{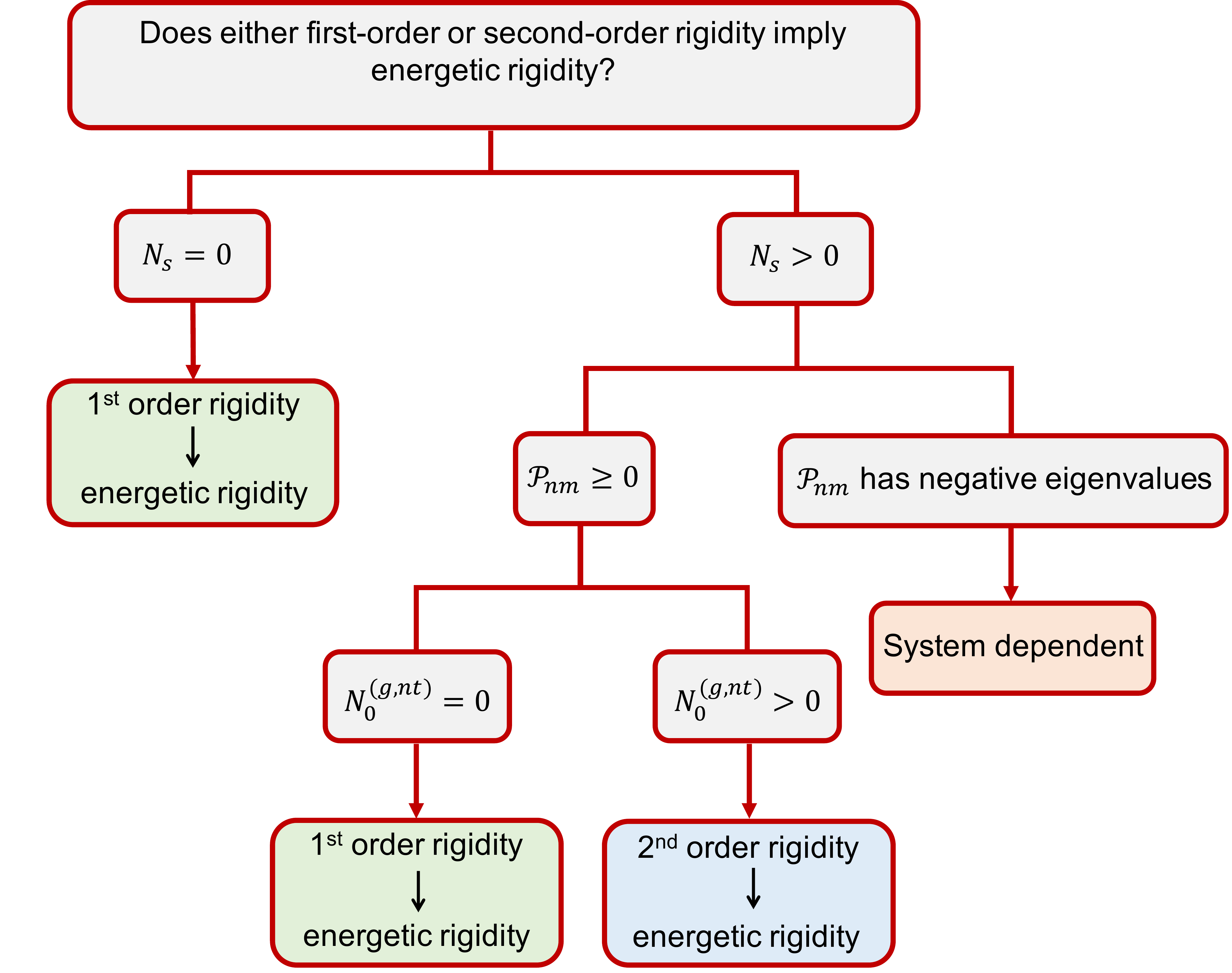}
    \caption{\textbf{Flowchart of Cases} summarizing the classification of systems based on the findings of second-order rigidity formalism. $N_0^{(g,nt)}$ refers to the number of global non-trivial LZMs (i.e.\ global FMs).}
    \label{fig:flowchart}
\end{figure}

Fig.~\ref{fig:flowchart} summarizes the cases describing when either first-order or second-order rigidity imply energetic rigidity. In Appendix~\ref{app:formalism}, we provide another flowchart (Fig.~\ref{relations}) to clearly establish the connection between energetic rigidity and structural rigidity as understood by mathematicians. We also provide several propositions to show that energetic rigidity and structural rigidity are interchangeable when $E=0$ but not necessarily otherwise. For instance, it can be shown that first-order and second-order rigidity both imply structural rigidity~\cite{Connelly2015}, but we saw that they do not always imply energetic rigidity. This is because for a system which possesses self stress at an energy minimum, mathematicians only require the existence of a linear combination of self stresses that would make the system rigid~\cite{Connelly1996}, however, that particular self stress may not be the linear combination of self stresses that the system chooses as its prestress based on external forces~\cite{Sartor2020}.

\section{Discussion and Conclusions}
We term an ``energetically rigid" structure as one where any sufficiently small applied displacement increases the structure's energy. Our focus on motions that preserve energy contrasts with previous work on structural rigidity that has focused on motions that preserve constraints. There are interesting differences between the two approaches. Unlike structural rigidity, energetic rigidity is not defined solely by the geometry -- predictions also depend on the energy functional. Here we studied a Hooke-like energy that is quadratic in the constraints, which is the simplest nontrivial energy functional that encompasses a large number of physical systems, but other choices are possible. On the other hand, this choice opens the possibility that in some structures there may exist motions that preserve the energy without preserving individual constraints. Importantly, the framework developed here would allow us to identify such systems as floppy.

Specifically, we want to understand under which precise circumstances structural rigidity implies energetic rigidity, and in the process identify underlying geometric mechanisms that are responsible for rigidity in specific materials. It is understood that predicting whether a planar graph is structurally rigid is already an NP-hard problem, and so previous work has proposed several ``quick" tests for rigidity, which work in limited circumstances. One test is the Maxwell-Calladine index theorem, also called first-order rigidity, which tests whether the constraints $f_\alpha$ that define the energy functional can be satisfied to first order. Another test is second-order rigidity, which checks whether constraints can be satisfied to second order.

In this work we have developed a systematic framework that clarifies the relationship between energetic rigidity and these other previously proposed rigidity tests. We demonstrate that first-order rigidity always implies energetic rigidity when there are no states of self stress. However, when the system does possess states of self stress, the eigenvalue spectrum of the prestress matrix $\mathcal{P}_{nm}$ controls whether first- or second-order rigidity (or neither) implies energetic rigidity. In a companion paper~\cite{damavandi2021b}, we study several physical systems of interest, and demonstrate that for some second-order rigidity is sufficient to guarantee energetic rigidity, while for others it is not. In particular, we use the formalism developed here to demonstrate that several important biological materials are second-order rigid and identify specific features of the eigenvalue spectrum and states of self stress, which drive biological processes, that arise due to second-order rigidity.

When the prestress matrix is indefinite or negative semi-definite, we can still show analytically that at the rigidity transition, second-order rigidity implies energetic rigidity. But away from the transition point neither first-order nor second-order rigidity guarantee energetic rigidity.

Moving forward, it would be useful to identify features that distinguish examples in this category, dividing it into sub-cases that are at least partially analytically tractable. One intriguing possibility is to classify a structure's response to applied loads.  For example, one could artificially increase the prestresses in a structure, multiplying $\mathcal{P}_{nm}$ by a coefficient $\epsilon>1$, which will only increase the overall magnitude of the state of self stress but not change the geometry of the network or the Gram term in the Hessian.   

This also suggests that it may be possible to program transitions between minima in the potential energy landscape via careful design of applied load. For example, while the type of spring network we study in our companion paper is completely tensile for $L_0<L_0^*$~\cite{damavandi2021b}, one could create spring networks with both tensile and compressed edges \cite{Bose2019} or a tensegrity with tensile cables and compressed rods. It will be interesting to see if we can design such systems to have a negative-definite prestress matrix. If so, applied loads may destabilize the structure along a specified mode towards a new stable configuration. These instabilities can also lead to more complex behaviors like dynamic snap-throughs, which can be identified using dynamic stability analyses \cite{Mascolo2019}. 

A related question is whether we can move such a system from one energy minimum to another in a more efficient manner. Traditionally, to push a system out of its local minimum into a nearby minimum, one rearranges the internal components of the system locally or globally, while it is rigid, by finding a saddle point on the energy landscape. An alternate design could be to (1) apply a global perturbation that makes the system floppy, (2) rearrange its components at no energy cost, and (3) apply a reverse global perturbation to make it rigid again.  In other words, the fact that the system can transition from rigid to floppy using very small external forces without adding or removing constraints could help us generate re-configurable materials with very low energy cost.

Another interesting avenue for design is to perturb the energy functional itself. In this work we focused on an energy that is Hookean in the constraints, but it would be interesting to explore whether different choices of energy functional still generate the same relationships between energetic rigidity and first- or second-order rigidity identified in Fig~\ref{fig:flowchart}. If not, such functionals may enable structures with interesting floppy modes.

Taken together, this highlights that the subtleties involved in determining energetic rigidity could be exploited to drive new ideas in material design.  With the framework described here, we now fully understand when we can use principles based on first-order constraint counting or second-order rigidity to ensure energetic rigidity in designed materials. Moreover, there may be some new design principles available, especially for dynamic and activated structures, if we focus on cases where these standard proxies fail. 

\begin{acknowledgments}
 We are grateful to Z. Rocklin for an inspiring initial conversation pointing out the connection between rigidity and origami, and to M. Holmes-Cerfon for substantial comments on the manuscript. This work is partially supported by grants from the Simons Foundation No 348126 to Sid Nagel (VH), No 454947 to MLM (OKD and MLM) and No 446222 (MLM). CDS acknowledges funding from the NSF through grant DMR-1822638, and MLM acknowledges support from NSF-DMR-1951921.
\end{acknowledgments}

\appendix
\begin{widetext}

\section{Derivation of second-order rigidity condition and implications for energetic rigidity}\label{app:formalism}

In Sec. \ref{sec:math}, we summarize the basic definitions and important theorems on structural rigidity in bar-joint frameworks. Several of these theorems are adapted from \cite{Connelly1996}. In Sec. \ref{sec:energeticrigidity}, we relate structural rigidity to energetic rigidity. These results are summarized in Fig.~\ref{relations}. We also provide derivations of second-order rigidity and energetic rigidity that we have omitted from the main text.

\subsection{Basic results on structural rigidity}\label{sec:math}

Let $x_n$ be a point in a space of configurations and let $\mathcal{F}_\alpha(\{x_n\})$ be a set of measures (for example, in a fiber network $\mathcal{F}_\alpha(\{x_n\})$ might give the length of the fibers). From now on we denote the configuration $\{x_n\}$ as $x$ for simplicity. We start with some basic definitions:\\

\mydef{A nontrivial isometry (or, sometimes, flex) is a one-parameter family of deformations, $x(t)$, such that $\mathcal{F}_\alpha(x(t)) = F_\alpha$ (for some $F_\alpha$) and $x(t)$ is not a translation or rotation. We similarly refer to a nontrivial deformation as any deformation $\delta x(t)$ that is not a translation or rotation.}

\mydef{A linear zero mode, also known as a first-order isometry or a first-order flex, at a configuration $\bar{x}$, $\dot{x}$, is a solution to the equation $\sum_n \partial_n \mathcal{F}_\alpha(\bar{x}) \dot{x}_n = 0$. A system is first-order rigid if there are no solutions to this equation.}

\mydef{A self stress, $\sigma_\alpha$, at $\bar{x}$ is a solution to $\sum_\alpha \sigma_\alpha \partial_n \mathcal{F}_\alpha(\bar{x}) = 0$.}

\mydef{A second-order isometry (or a second-order flex) at $\bar{x}$ is a first-order isometry such that $\sum_\alpha \sum_{nm} \sigma_{\alpha,I} \partial_n \partial_m \mathcal{F}_\alpha(\bar{x}) \dot{x}_n \dot{x}_m= 0$ has a solution where $ \{ \sigma_{\alpha,1}, \sigma_{\alpha,2}, \cdots, \sigma_{\alpha,N_s} \}$ is a basis of self stresses at $\bar{x}$. A system is second-order rigid if it has nontrivial zero modes but no nontrivial second-order isometries.}

We finally have a main result of rigidity theory: \emph{a system that is either first-order or second-order rigid, is structurally rigid} \cite{Connelly1996}. It can be hard -- still -- to test for structural rigidity at second order because it involves solving a system of quadratic equations. It is, therefore, convenient to introduce a stronger condition:

\mydef{A system is prestress stable at $\bar{x}$ if there is a self stress at $\bar{x}$, $\sigma_\alpha$, such that $\sum_\alpha \sigma_\alpha \partial_n \partial_m \mathcal{F}_\alpha(\bar{x})$ is positive definite on every nontrivial zero mode.}

With this definition, we prove that \emph{a system that is prestress stable at $\bar{x}$ is also second-order rigid at $\bar{x}$ (and hence, structurally rigid)}. This follows because there is a self stress $\sigma_\alpha$ such that $\sum_\alpha \sigma_\alpha \partial_i \partial_j \mathcal{F}_\alpha(\bar{x})$ is positive definite on nontrivial first-order flexes. We can construct a basis for the self stresses with $\sigma_\alpha$ as one of its elements. Therefore, it is second-order rigid as well.

According to Connelly and Whitely~\cite{Connelly1996}, there are examples of second-order rigid structures that are not prestress stable in 2D and, especially, 3D. The notion of prestress stability is related to notions of an energy.

Note also that a system that is second-order rigid is not necessarily prestress stable. Examples appear in Connelly and Whitely. However,

\prop{A system that is second-order rigid but has one self stress is prestress stable. This is also in~\cite{Connelly1996}.}

We must have $c \sigma_\alpha \partial_n \partial_m f(\bar{x})$ positive definite for some, potentially negative, $c$. Then choosing $F_\alpha = \mathcal{F}_\alpha(\bar{x}) - c \sigma_\alpha$ is energetically rigid to quadratic order and, hence, prestress stable.

\begin{figure}
\centering
\includegraphics[width=0.8 \linewidth]{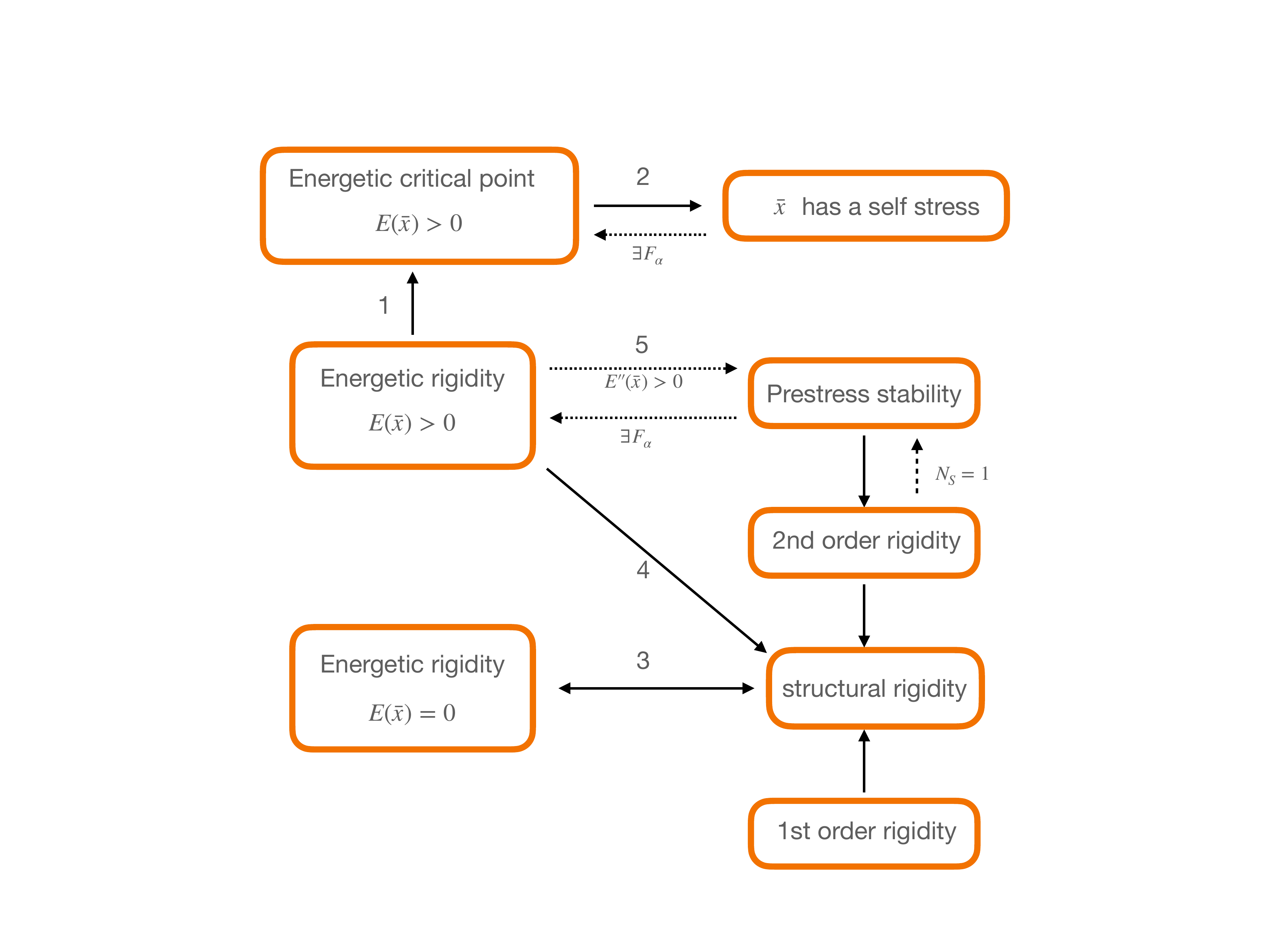}
\caption{\textbf{Relations between various definitions} for a given configuration $\bar{x}$. The numbers on arrows refers to propositions with the same numbers. We can see that only when the system is unstressed ($E(\bar{x})=0$), energetic rigidity and structural rigidity are equivalent (one is always guaranteed to imply the other). Dotted arrows labeled with $\exists F_\alpha$ mean that the implication is only valid for specific choices of $F_\alpha$ and thus prestress. $E''(\bar{x})>0$ denotes energetic rigidity at quadratic order (positive-definite Hessian). Dashed arrow with $N_s=1$ means that the implication is guaranteed when there is only one state of self stress.}
\label{relations}
\end{figure}

\subsubsection{Energetic rigidity}\label{sec:energeticrigidity}

A proper understanding of the rigidity of a mechanical system requires an energy functional. To formulate this, we assume we have a system of measures, $\mathcal{F}_\alpha( \left\{ x \right\} )$. From this we define generalized strains, $f_\alpha(x) = \sqrt{k_\alpha} \left( \mathcal{F}_\alpha(x)-F_\alpha\right)$ that measure the deformation of our system from the local equilibrium $F_\alpha$ and $k_\alpha > 0$ is an elastic modulus. We then assume a neo-Hookean energy functional of the form
\begin{equation}\label{eq:E}
    E(x) = \frac{1}{2} \sum_\alpha f_\alpha^2(x).
\end{equation}
As an example, for a fiber network, $\mathcal{F}_\alpha(x)$ measures the distance between two vertices and $F_\alpha$ is the equilibrium distance between vertices. For a vertex model, on the other hand, the $f_\alpha$ might measure the deviation of the cell perimeters and areas from their equilibrium values.

We say that \emph{a system is energetically rigid at $\bar{x}$ if there exists a $c$ such that $E( \bar{x} + \epsilon \delta x ) > E(\bar{x})$ for any nontrivial deformation $\delta x$ and any $0 < \epsilon < c$.} In other words, it is energetically rigid if all sufficiently small, finite deformations increase the energy. This conforms to the intuitive notion that a system is rigid if deforming it increases the energy. Similarly, \emph{a system is energetically rigid at $n^{th}$ order at the configuration $\bar{x}$ if $\sum_{i_1 \cdots i_n}\partial_{i_1} \cdots \partial_{i_n} E (\bar{x}) \delta x_{i_1} \cdots \delta x_{i_n} > 0$ for any nontrivial deformation, $\delta x$}.

Unsurprisingly, the notion of energetic rigidity is closely allied with structural rigidity and its various proxies. These notions are, however, not identical, and here we discuss the many interconnections between structural and energetic rigidity. These relationships are summarized in Fig.~\ref{relations}. Important to note is that the dashed arrows signify that while the implication can be proved for some choice of self stress, it is not guaranteed that a given system has picked that particular self stress at the energy minimum (i.e.\ the actual prestress may be a different linear combination of self stresses). The numbers labeling the propositions below refer to the arrows in Fig.~\ref{relations} labeled with the same number.

\prop{(1) Energetic rigidity at $\bar{x}$ with $E(\bar{x}) > 0$ implies $\bar{x}$ is a critical point of the energy.}
Let $\bar{x}$ be a point that is energetically rigid. This means that $E(\bar{x} + \epsilon \delta x) > E(\bar{x})$ for all nontrivial $\delta x$ and for all $0 < \epsilon < c$. Taking the derivative with respect to $\epsilon$ gives
\begin{equation}
\lim_{\epsilon \rightarrow 0+} \partial_\epsilon E(\bar{x}+ \epsilon \delta x) = \sum_n \partial_n E(\bar{x}) \delta x_n.
\end{equation}
If this were not a critical point then taking $\delta x \rightarrow - \delta x$ would give us a nontrivial deformation that decreases the energy for some $\epsilon$ that was small enough. Therefore, it must be a critical point.

\prop{(2) The point $\bar{x}$ is a critical point of some energy with $E(\bar{x}) > 0$ if there is a self stress at $\bar{x}$. The converse is also true for specific choices of $F_\alpha$.}
We first assume $\bar{x}$ is a critical point with $E(\bar{x}) > 0$. Then $\partial_n E( \bar{x}) = 0$, which requires
\begin{equation}
0 = \sum_\alpha \left[ \mathcal{F}_\alpha(\bar{x}) - F_\alpha \right] \partial_n \mathcal{F}_\alpha(\bar{x}).
\end{equation}
Since $E(\bar{x}) \ne 0$, $\mathcal{F}_\alpha(\bar{x}) \ne F_\alpha$. Therefore, $\mathcal{F}_\alpha(\bar{x}) - F_\alpha$ is a self stress.

Now assume that we have a point $\bar{x}$ where $\sigma_\alpha$ is a self stress. Then choose $F_\alpha = \mathcal{F}_\alpha(\bar{x}) + c \sigma_\alpha$. We can now verify that $\bar{x}$ is a critical point of $E(x) = \sum_\alpha [ \mathcal{F}_\alpha(x) - \mathcal{F}_\alpha(\bar{x}) + c \sigma_\alpha ]^2$ for any $c$.

\prop{(3) The configuration $\bar{x}$ is energetically rigid at $E(x)$ with $E(\bar{x}) = 0$ if and only if $\bar{x}$ is structurally rigid.}
We first assume that $\bar{x}$ is structurally rigid. Then let $F_\alpha = \mathcal{F}_\alpha(\bar{x})$. We get $E(\bar{x}) = 0$. Let $\delta x$ be any nontrivial deformation. Since $\mathcal{F}_\alpha(\bar{x} + c \delta x) \ne F_\alpha$ for sufficiently small $c$ we must have $E(\bar{x}+c \delta x) > 0$ implying the system is energetically rigid.

Now assume we have an energy such that $\bar{x}$ is energetically rigid with $E(\bar{x}) = 0$. Then $\mathcal{F}_\alpha(\bar{x}) = F_\alpha$. Since $E(\bar{x}+c \delta x) > 0$ for appropriately chosen $c$, we must have $\mathcal{F}_\alpha(\bar{x} + c \delta u) \ne F_\alpha$.

\prop{(4) Let $\bar{x}$ be an extremum of $E(x)$ such that $E(\bar{x}) \ne 0$ and suppose that $\bar{x}$ is energetically rigid. Then the system is structurally rigid at $\bar{x}$ as well.}
Suppose that $\bar{x}$ is an extremum of $E(x)$ such that $E(\bar{x}) \ne 0$ but such that $\bar{x}$ is energetically rigid. That is, all nontrivial directions raise the energy further. Then there cannot be any nontrivial isometries $x(t)$ passing through $\bar{x}$ since if there were $E$ would have to be constant along them and this contradicts the assumption.

Note that this can be extended to energy maxima as well. The converse need not be true though. If a system is rigid at $\bar{x}$, choosing $F_\alpha$ so that $\bar{x}$ is an extremum does not mean that it will be energetically rigid. Let's suppose that $x(t)$ is a one-parameter family of constant energy trajectories. Then
\begin{equation}
\partial_t E[ x(t) ] = 0 = \sum_\alpha\sum_n [\mathcal{F}_\alpha(x(t)) - F_\alpha] \partial_n \mathcal{F}( x(t) ) \dot{x}_n.
\end{equation}
This can only be true if $x(t)$ are all extrema of $E$ with $E(x(t)) \ne 0$. In addition, there must be at least one self stress along the entire trajectory $x(t)$.

The notion of prestress stability is intimately related to energetic rigidity at quadratic order. The next proposition establishes the equivalence of prestress stability (as defined above) and energetic rigidity to quadratic order:

\prop{(5) A system is prestress stable at $\bar{x}$ if and only if there is a choice $F_\alpha$ such that it is an extremum of the energy with $E(\bar{x}) \ne 0$ and is energetically rigid at quadratic order.}

To prove this we first assume that the system is prestress stable and let $\sigma_\alpha$ be the self stress such that $\sum_\alpha \sigma_\alpha \partial_n \partial_m \mathcal{F}_\alpha(\bar{x})$ is positive definite on nontrivial first-order flexes. Then define an energy functional
\begin{equation}
E(x) = \sum_\alpha \left[ \mathcal{F}_\alpha(x) - \mathcal{F}_\alpha(\bar{x}) + c \sigma_\alpha \right]^2,
\end{equation}
where $c > 0$ is some arbitrary number. We can now check that $\bar{x}$ is an extremum, $\partial_n E(\bar{x}) = c \sum_\alpha \sigma_\alpha \partial_n \mathcal{F}_\alpha(\bar{x}) = 0$. Computing the Hessian, we find
\begin{equation}
H_{n m} = \sum_\alpha \partial_n \mathcal{F}_\alpha(\bar{x}) \partial_m \mathcal{F}_\alpha(\bar{x}) + c \sum_\alpha \sigma_\alpha \partial_n \partial_m \mathcal{F}_\alpha(\bar{x}).
\end{equation}
This is positive definite on nontrivial first-order flexes by the assumption of prestress stability, for any $c$. On modes that are not nontrivial first-order flexes, we can always choose $c > 0$ sufficiently small that the first term dominates (choose $c$ to be smaller than the smallest eigenvalue of the Gram term). Therefore, $\bar{x}$ is an energetically stable extremum of $E(x)$ when $F_\alpha = f_\alpha(\bar{x}) - c \sigma_\alpha$.

Going the other way, let's assume that our system is energetically rigid at quadratic order at an extremum $\bar{x}$. Then let $\dot{x}_n$ be any nontrivial, first-order flex. We have
\begin{equation}
\sum_{nm} H_{nm} \dot{x}_n \dot{x}_m = \sum_{nm}\sum_\alpha [\mathcal{F}_\alpha(\bar{x}) - F_\alpha] \partial_n \mathcal{F}(\bar{x}) \dot{x}_n \dot{x}_m > 0.
\end{equation}
That implies that $\mathcal{F}_\alpha(\bar{x}) - F_\alpha$ is a self stress and that it is prestress stable.

It is worth noting that prestress stability at $\bar{x}$ does not imply that a system is energetically rigid at $\bar{x}$ for a particular choice of $F_\alpha$, only for some choice.

We have already seen that second-order rigidity does not imply prestress stability in the last section. Here we note that prestress stability and energetic rigidity are not identical either. In particular, a system that is prestress stable may not be energetically rigid for a particular choice of $F_\alpha$. Suppose that a system is prestress stable but has a self stress $\sigma_\alpha$ for which the prestress matrix is not positive definite on the nontrivial first-order flexes. Choose $F_\alpha = \mathcal{F}_\alpha(\bar{x}) - c \sigma_\alpha$. This shows that the system with this choice is not energetically rigid at quadratic order. In other words, the prestress that the system picks at $\bar{x}$ may not be one that makes the system prestress stable. If there is only one self stress and the system is prestress stable, then energetic rigidity and prestress stability trivially imply each other.

Finally, the following proposition deals with the nonlinear nature of rigidity:

\prop{A system is energetically rigid at $\bar{x}$ with $E(\bar{x}) = 0$ to fourth order if it is second-order rigid.}

This proposition shows that even if the standard checks of energetic rigidity (e.g.\ shear modulus) suggest floppiness, the system may still be energetically rigid to finite deformations. We will prove this proposition in the following section, where we also show a more detailed derivation of the equations in section~\ref{sec:formalism}. All of these results demonstrate that the relationships between all of these notions of rigidity are, in fact, quite subtle.

\subsection{Second-order rigidity and energetic rigidity}
Our goal here is to derive conditions for second-order zero modes and study the energy of systems that are second-order rigid. We will show that a system that has no prestress (Case 2A) but is second-order rigid is energetically rigid as well at fourth order in deformations. For prestressed systems, we show derivations of our claims for Case 2B and 2C.

Take constraints $f_\alpha$ on a given system, e.g., $f_\alpha(\{x_n\})$ may be the displacements of edges of a graph from their equilibrium lengths. The energy functional is $E = k\sum_{\alpha=1}^M f_\alpha^2/2$ where $M$ is the number of constraints. We set $k=1$ without loss of generality. For a more general case with constraint dependent stiffnesses $k_\alpha$, we can simply re-scale the constraints to $f'_\alpha = \sqrt{k_\alpha}f_\alpha$.  Imagine that $\bar{x}_n$ is at a critical point of $E$. 

At a critical point, $\sum_\alpha f_\alpha(\{\bar{x}_n\}) \partial_m f_\alpha(\{\bar{x}_n\}) = 0$. Let $\{ \sigma_{\alpha,1},\cdots , \sigma_{\alpha,N_s}, e_{\alpha,1}, \cdots , e_{\alpha,M-N_s} \}$ be an orthonormal basis in $\mathbb{R}^M$ where $\sum_\alpha \sigma_{\alpha,I} \cdot \partial_n f_\alpha(\{\bar{x}_n\}) = 0$ (so $\sigma_{\alpha,I}$ are self stresses).
Let us further assume $f_\alpha(\{\bar{x}_n\}) = C \sigma_{\alpha,1}$ with $C > 0$, which we can do without loss of any generality.

To find zero modes, we Taylor expand $f_\alpha$ for small perturbations around $\bar{x}_n$. To easily keep track of the order of expansion, we parametrize deformations in time so that at an infinitesimal time $\delta t$ we have a deformation $x_n(\delta t)$ such that $x_n(0) = \bar{x}_n$. We then have
\begin{equation}
f_\alpha(\{x_n(\delta t)\}) \approx C \sigma_{\alpha,1} + \sum_n \partial_n f_\alpha \dot{x}_n \delta t +\frac{1}{2}\left[ \sum_n \partial_n f_\alpha \ddot{x}_n+\sum_{nm} \partial_n \partial_m f_\alpha \dot{x}_n \dot{x}_m\right]\delta t^2 + \mathcal{O}(\delta t^3),
\end{equation}
where partial derivatives are evaluated at $\bar{x}_n$. Also, $\dot{x}_n$ is short hand for $\dot{x}_n(0)$ and $\ddot{x}_n$ is short hand for $\ddot{x}_n(0)$. That is, these are explicitly independent vectors that determine the first two terms in a Taylor expansion of $x_n(t)$ around $t=0$.

It is useful to project $f_\alpha(\{x_n(\delta t)\})$ along the orthonormal basis vectors

\begin{align} \label{eq:constraint1}
&\sum_{\alpha} \sigma_{\alpha,I} f_\alpha(\{x_n(\delta t)\}) \approx C \delta_{I 1} + \sum_{\alpha}\sum_{nm} \sigma_{\alpha,I} \partial_n \partial_m f_\alpha \dot{x}_n \dot{x}_m  \delta t^2, \\ \label{eq:constraint2}
&\sum_{\alpha} e_{\alpha,I} f_\alpha(\{x_n(\delta t)\}) \approx \sum_{\alpha}e_{\alpha,I} \sum_n \partial_n f_\alpha \dot{x}_n \delta t + \frac{1}{2} \sum_{\alpha} e_{\alpha,I} \left[ \sum_n \partial_n f_\alpha \ddot{x}_n+\sum_{nm} \partial_n \partial_m f_\alpha \dot{x}_n \dot{x}_m\right] \delta t^2.
\end{align}

To find second-order zero modes, modes that preserve $f_\alpha$ to second order, Eqs.~(\ref{eq:constraint1}-\ref{eq:constraint2}) imply the system
\begin{eqnarray}
\sum_{\alpha}e_{\alpha,I} \sum_n \partial_n f_\alpha \dot{x}_n &=& 0 \nonumber \\
\sum_{\alpha} e_{\alpha,I} \left[ \sum_n \partial_n f_\alpha \ddot{x}_n+\sum_{nm} \partial_n \partial_m f_\alpha \dot{x}_n \dot{x}_m\right] &=& 0 \nonumber \\
\sum_{\alpha}\sum_{nm} \sigma_{\alpha,I} \partial_n \partial_m f_\alpha \dot{x}_n \dot{x}_m  &=& 0 \nonumber
\end{eqnarray}
where the first equation implies $\dot{x}_n$ is along a linear zero mode (note that $\sum_n \partial_n f_\alpha \dot{x}_n$ must have a non-zero projection on at least one $e_{\alpha,I}$ since it is perpendicular to all self stresses $\sigma_{\alpha,I}$ by definition),  the middle equation is associated to the \textit{curvature} of the linear zero mode as we proceed along $t$, and the last equation gives an additional quadratic constraint that these tangents must satisfy to be second-order zero modes. Multiplying the last equation by $\delta t^2$, we recover Eq.~(\ref{eq:secondorder}).

Notice that the middle equation always has a solution. To see this, we note that it is a linear equation of the form $A \ddot{x} - b = 0$. Since $b$ is explicitly in the image of $A$, $\ddot{x}$ has a solution that is unique up to zero modes. Since the linear zero modes are already included in $\dot{x}_n$, we can choose $\ddot{x}_n$ to be orthogonal to them without loss of generality. With that choice, the matrix $\sum_{\alpha}e_{\alpha,I} \partial_n f_\alpha$ is invertible.

Putting all of this into the energy, we find that

\begin{align}
E \approx \frac{1}{2} \sum_{I=1}^{M-N_s} \left[ \sum_{\alpha}e_{\alpha,I} \sum_n \partial_n f_\alpha \dot{x}_n + \frac{1}{2} \sum_{\alpha} e_{\alpha,I} \left[ \sum_n \partial_n f_\alpha \ddot{x}_n+\sum_{nm} \partial_n \partial_m f_\alpha \dot{x}_n \dot{x}_m\right] \delta t \right]^2 \delta t^2 \\
 +\frac{1}{2} \left[ C + \frac{1}{2} \sum_{\alpha}\sum_{nm} \sigma_{\alpha,1} \partial_n \partial_m f_\alpha \dot{x}_n \dot{x}_m  \delta t^2 \right]^2 + \frac{1}{8} \sum_{I=2}^{N_s} \left[  \sum_{\alpha}\sum_{nm} \sigma_{\alpha,I} \partial_n \partial_m f_\alpha \dot{x}_n \dot{x}_m \right]^2  \delta t^4.\nonumber
\end{align}

What we are interested in is whether we can find a solution $x_n(t)$ such that $E(t)$ increases, decreases, or stays constant to a particular order in $\delta t$.

Let us consider what happens when $C \rightarrow 0$ first. Note that some systems may not be able to achieve a state with $C=0$ because of the way they are prepared. Here, we assume that the energy can be continuously modulated to zero. Such a system is not prestressed, but can still possess self stresses (e.g.\ the onset of geometric incompatibility \cite{Merkel2019}).  In that case,

\begin{align}
E &\approx \frac{1}{2} \sum_{I=1}^{M-N_s} \left[ \sum_{\alpha}e_{\alpha,I} \sum_n \partial_n f_\alpha \dot{x}_n + \frac{1}{2} \sum_{\alpha} e_{\alpha,I} \left[ \sum_n \partial_n f_\alpha \ddot{x}_n+\sum_{nm} \partial_n \partial_m f_\alpha \dot{x}_n \dot{x}_m\right] \delta t \right]^2 \delta t^2 \\
&+ \frac{1}{8} \sum_{I=1}^{N_s} \left[  \sum_{\alpha}\sum_{nm} \sigma_{\alpha,I} \partial_n \partial_m f_\alpha \dot{x}_n \dot{x}_m \right]^2  \delta t^4.\nonumber
\end{align}

The energy is constant as long as the coefficients of $\delta t^2$, $\delta t^3$, and so on vanish. These lead to
\begin{equation}\label{eq:udot}
\sum_{\alpha}e_{\alpha,I} \sum_n \partial_n f_\alpha \dot{x}_n = 0,
\end{equation}
to second order, and we have the two equations
\begin{equation}\label{eq:udoubledot}
\sum_{\alpha} e_{\alpha,I} \left[ \sum_n \partial_n f_\alpha \ddot{x}_n+\sum_{nm} \partial_n \partial_m f_\alpha \dot{x}_n \dot{x}_m\right] = 0,
\end{equation}
and
\begin{equation}\label{eq:quadratic}
\sum_{\alpha}\sum_{nm} \sigma_{\alpha,I} \partial_n \partial_m f_\alpha \dot{x}_n \dot{x}_m  = 0, 
\end{equation}
to fourth order. The third order term already vanishes if the quadratic term vanishes. These are the three equations that defined a quadratic isometry previously. Hence, $E$ is constant along any quadratic isometry. Similarly, if $E$ is constant along a direction, the trajectory must be along a quadratic isometry. So at the critical point, second-order rigidity implies energetic rigidity to this order in $\delta t$. This also proves the last proposition in the previous section.

Now, one might wonder what happens as $C$ increases. We then have
\begin{align}
E &= \frac{C^2}{2} + \frac{1}{2} \delta t^2 \left[ \sum_{I=1}^{M-N_s} \left(\sum_\alpha e_{\alpha,I} \sum_n \partial_n f_\alpha \dot{x}_n\right)^2  +  \sum_{\alpha}\sum_{nm} C \sigma_{\alpha,1} \partial_n \partial_m f_\alpha \dot{x}_n \dot{x}_m \right] \nonumber \\
& + \frac{1}{2} \delta t^3 \sum_{I=1}^{M-N_s} \left(\sum_\alpha e_{\alpha,I} \sum_n \partial_n f_\alpha ~ \dot{x}_n\right) \left(\sum_{\alpha'} e_{\alpha',I} \left[ \sum_n \partial_n f_{\alpha'} \ddot{x}_n+\sum_{nm} \partial_n \partial_m f_{\alpha'} \dot{x}_n \dot{x}_m\right]\right) \\
& +  \frac{1}{8} \delta t^4 \sum_{I=1}^{M-N_s} \left(\sum_{\alpha} e_{\alpha,I} \left[ \sum_n \partial_n f_\alpha \ddot{x}_n+\sum_{nm} \partial_n \partial_m f_\alpha \dot{x}_n \dot{x}_m\right]\right)^2   + \frac{1}{8} \delta t^4 \sum_{I=1}^{N_s} \left[\sum_{\alpha}\sum_{nm} \sigma_{\alpha,I} \partial_n \partial_m f_\alpha \dot{x}_n \dot{x}_m \right]^2. \nonumber
\end{align}

The second-order term is the Hessian. If that has a direction that is negative, then we have not expanded around a local minimum. However, one can ask whether or not zero directions might arise even if the system is second-order rigid. For that to happen, however, $\dot{x}_n$ cannot be along a zero mode. If it was along a zero mode and the Hessian was zero, the fact that the system is second-order rigid would imply that the energy increases to fourth order. If $\dot{x}_n$ was not along a zero mode and the Hessian was zero, for it to not increase the energy to the fourth order, it has to satisfy Eq.~(\ref{eq:quadratic}), similar to second-order zero modes (this system would belong to Case 2C). 
\end{widetext}

\bibliography{apssamp}

\end{document}